# The Heats of Reactions. Calorimetry and Van't-Hoff. 1

I. A. Stepanov

Latvian University, Rainis bulv. 19, Riga, LV-1586, Latvia

## Abstract

Earlier it has been supposed that the law of conservation of energy in chemical reactions has the following form:

$$dU = dQ - PdV + \sum_i \mu_i dN_i$$

In [1-4] it has been shown that for the biggest part of reactions it must have the following form:

$$dU = dQ + PdV + \sum_i \mu_i dN_i$$

In the present paper this result is confirmed by other experiments.

## 1. Introduction

For chemical processes the law of conservation of energy is written in the following form:

$$dU = dQ - PdV + \sum_i \mu_i dN_i \qquad (1)$$

where dQ is the heat of reaction, dU is the change in the internal energy, $\mu_i$ are chemical potentials and $dN_i$ are the changes in the number of moles.

In [1-4] it has been shown that the energy balance in the form of (1) for the biggest part of the chemical reactions is not correct. In the biggest part of the chemical reactions the law of conservation of energy must have the following form:

$$dU=dQ+PdV+\sum_i \mu_i dN_i \qquad (2)$$

The Van't-Hoff equation is the following one:

$$d/dT \ln K = \Delta H^0/RT^2 \qquad (3)$$

where K is the reaction equilibrium constant and $\Delta H^0$ is the enthalpy. According to thermodynamics, the Van't-Hoff equation must give the same results as calorimetry because it is derived from the 1st and the 2nd law of thermodynamics without simplifications. However, there is a paradox: the heat of chemical reactions, that of dilution of liquids and that of other chemical processes measured by calorimetry and by the Van't-Hoff equation differ significantly [1-4]. The difference is far beyond the error limits. The reason is that in the derivation of the Van't-Hoff equation it is necessary to take into account the law of conservation in the form of (2), not of (1) [1-4].

If to derive the Van't-Hoff equation using (2) the result will be the following one:

$$d/dT \ln K = \Delta H^{0*}/RT^2 \qquad (4)$$

where $\Delta H^{0*} = \Delta Q^0 + P\Delta V^0$.

In the present paper the following processes were checked: Si(liquid)=Si(gas), 2Si(liquid)=Si$_2$(gas), Si(solid)+SiO$_2$(solid)=2SiO(gas).

## 2. Experiments

In [6] the following processes were considered:

$$Si(liquid)=Si(gas) \tag{5}$$

$$\log P = \log K = -2{,}08 \cdot 10^4/T + 10{,}84 \qquad 1762 \leq T \leq 1998K \tag{6}$$

$$2Si(liquid)=Si_2(gas) \tag{7}$$

$$\log P = \log K = -2{,}46 \cdot 10^4/T + 10{,}93 \qquad 1842 \leq T \leq 1992K \tag{8}$$

The 1st process is evaporation, the 2nd one is a chemical reaction. Their heats are given in **Table 1**. One sees that the heat of evaporation measured by the Van't-Hoff equation is much closer to the experiment than the heat of the reaction. Evaporation is a physical process, it obeys the traditional thermodynamics. The second process obeys the present theory.

In [7] the following reaction has been studied:

$$Si(solid)+SiO_2(solid)=2SiO(gas) \tag{9}$$

$$\log P = 13{,}613 - 1{,}785 \cdot 10^4/T, \quad 1300 \leq T \leq 1580K \tag{10}$$

The heat of this reaction is given in **Table 1**. It obeys the present theory, not the traditional one.

**Table 1**

The heat of evaporation and of some chemical reactions measured by the Van't-Hoff equation and by calorimetry

| T,°K | $\Delta H^0$, kJ/mol | $\Delta H^0-P\Delta V^0$, kJ/mol | $\Delta Q^0$, kJ/mol [5, 8] |
|---|---|---|---|
| Si(liquid)=Si(gas)[6]1 | | | |
| 1800 | 397,55 | - | 395,78 |
| 1900 | 397,55 | - | 395,23 |
| 2Si(liquid)=Si$_2$(gas)[6]1 | | | |
| T,°K | $\Delta H^{*0}$, kJ/mol | $\Delta H^{*0}-P\Delta V^0$, kJ/mol | $\Delta Q^0$, kJ/mol [5, 8] |
| 1850 | 470,19 | 454,82 | 455,57 |
| 1900 | 470,19 | 454,40 | 455,05 |
| 1990 | 470,19 | 453,65 | 454,11 |
| Si(solid)+SiO$_2$(solid)=2SiO(gas)[7] | | | |
| T,°K | $\Delta H^{*0}$, kJ/mol | $\Delta H^{0*}-P\Delta V^0$, kJ/mol | $\Delta Q^0$, kJ/mol[9] |
| 1300 | 683,10 | 661,50 | 661,81 |
| 1400 | 683,10 | 659,83 | 657,90 |
| 1500 | 683,10 | 658,17 | 652,43 |

---

[1] Thermodynamic data for Si(liquid) is taken from [5], these for Si(gas) and Si$_2$(gas) is from [8].